\documentclass[aps,prb,twocolumn,reprint,groupedaddress,showpacs]{revtex4-1}

\usepackage{graphicx}

\def\bM{\mathbf{M}}

\def\bz{\mathbf{z}}
\def\bH{\mathbf{H}}
\def\bx{\mathbf{x}}

\def\bxper{\mathbf{x}_{\bot}}

\def\bunitx{\hat{\mathbf{x}}}
\def\bunity{\hat{\mathbf{y}}}
\def\bunitz{\hat{\mathbf{z}}}

\def\bunitrho{\hat{\mathbf{\rho}}}
\def\bunitphi{\hat{\mathbf{\varphi}}}

\def\bk{\mathbf{k}}

\def\bq{\mathbf{q}}

\def\bkper{\mathbf{k}_{\bot}}
\def\bn{\mathbf{n}}
\def\bunitn{\hat{\mathbf{n}}}

\def\ml{m_{//}}

\def\barlH{\bar{l}_H}

\begin{document}


\title{Skyrmions and Their Sizes in Helimagnets}


\author{Kwan-yuet Ho}
\email[]{kwyho@umd.edu}
\author{Theodore R. Kirkpatrick}
\affiliation{Institute for Physical Science
  and Technology and Department of Physics, University of Maryland,
  College Park, MD 20742, USA}
\author{Dietrich Belitz}
\affiliation{Department of Physics and Institute of Theoretical
  Science, University of Oregon, Eugene, Oregon 97403, USA}


\date{\today}

\begin{abstract}
Skyrmion gases and lattices in helimagnets are studied, and the size
of a Skyrmion in various phases
is estimated. For isolated Skyrmions, the long distance tail is
related to the magnetization correlation functions and exhibits power-law
decay if the phase spontaneously breaks a continuous symmetry, but
decays exponentially otherwise. The size of a Skyrmion is
found to depend on a number of length scales. These
length scales are related to
the strength of Dzyaloshinskii-Moriya (DM) interaction, the thermal
correlation lengths, and the strength of the external magnetic
field. An Abrikosov lattice of Skyrmions is found to exist near the
helimagnetic phase boundary, and the core-to-core distance is estimated.
\end{abstract}

\pacs{12.39.Dc, 75.10.Hk, 75.50.-y}

\maketitle

\section{Introduction}

\begin{figure}[htp]
\centering
\includegraphics[scale=.98]{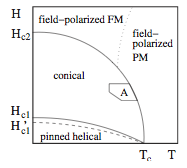}
\caption{Schematic phase diagram of {MnSi} in the $H$-$T$ plane
  showing the helical, conical, and $A$ phases, as well as the
  field-polarized ferromagnetic (FM) and paramagnetic (PM) states. \cite{HoKiSaBe10}} \label{gr_phasediagram_HoKiSaBe10}
\end{figure}

Skyrmions are non-trivial topological objects in various field
theories. They were first used to model baryons in nuclear physics. \cite{Sky61}
More recently, they have been observed in quantum Hall ferromagnets, \cite{FBCMS97,TiGiFe98}
$p$-wave superconductors, \cite{LiToBe09} and
Bose-Einstein condensates. \cite{Ho98, LHWDB09} The Skyrmion lattice
is also a candidate for
an experimentally observed periodic phase in helimagnets such as {MnSi} 
\cite{TPSF97,MBJPRNGB09,PNMJJLRMFNKGBBR09} (called the A phase)
and {Fe${}_{1-x}$Co${}_x$Si} (called the SkX phase). \cite{YOKPHMNT10} The
schematic phase diagram of a helimagnet is shown in Fig. \ref{gr_phasediagram_HoKiSaBe10}. In
these helimagnets, the usual ordered phase that exhibits long-range order
takes the form of a helix, \cite{BakJen80, PluWal81} that is
thermodynamically stabilized by Dzyaloshinskii-Moriya (DM) interaction.
\cite{Dzy58, Mor60}
The helix of the magnet can be characterized by the pitch vector
$\bq$, with magnitude being the helical wavenumber and the direction
being the direction of the helix. Its magnitude is proportional to the
strength of the DM interaction. The phase that is believed to be a
lattice of Skyrmions in these helimagnets have lattice size of the
order of magnitude $q^{-1}$. It is likely
that dilute Skyrmion gases in various phases of helimagnets can also be
stabilized through the DM interaction.

An isolated Skyrmion is an exact solution of the saddle-point equation
to the non-linear $\sigma$ model. \cite{BelPol75, AbaPok98}
Thermodynamically it is a metastable state. Different schemes have been proposed to
stabilize the Skyrmions, with a 
dipole-dipole interaction \cite{Eza10} and a DM interaction
\cite{BogYab89,BogHub94,Bog95,RoBoPf06,BLRB10,RoLeBo11}
being two examples. These schemes introduce new physical length
scales. 

A lattice of Skyrmions has also been described as the
superposition of three perpendicular
helimagnets. \cite{MBJPRNGB09,EGDR11} This Fourier description was argued to be a
lattice of Skyrmion by calculating the Skyrmion density. However, this
approximation is not a solution of the saddle-point equation to the
model they are using. \cite{HoKiSaBe10} Using the technique of Abrikosov vortices in type-II
superconductors, \cite{Abr57} a lattice of Skyrmions can
be described. \cite{HZYPN10} The Goldstone modes for such lattice
is the same as that of the columnar phase of the liquid
crystals. \cite{KirBel10} Some previous studies \cite{Bog95,BogHub94}
showed that by increasing the
magnetic field, the system changes from aligned conical phase, then a
Skyrmion lattice and finally to a ferromagnet. This is true for
certain temperatures but not for all. The Skyrmion lattice is likely
formed by a first-order phase transition.

The sizes of Skyrmions can be characterized in four ways. It
can be characterized by the behavior near the core. \cite{RoLeBo11}
Alternatively, the size can
also be characterized by the decay length of the long-distance tail
of the Skyrmion. In this paper, we introduce yet another measure of
the size of a of Skyrmion, which is given
by the length at which the behaviors of the core and the tail
match. We believe that it is the best measure of the size of a
Skyrmion. Finally, in a Skyrmion lattice, there is a fourth measure of
the size of a Skyrmion: the distance between cores in the lattice.

In this paper, we study both an isolated Skyrmion and a Skyrmion lattice in
the presence of DM interaction and external magnetic field, and we
estimate their sizes. In Sec. \ref{sec_model}, we
review the Landau-Ginzburg-Wilson (LGW) model with the DM interaction
used to study the helimagnets. Crucially, we also introduce the
various length scales in the problem. In
Sec. \ref{sec_basicsSkyr}, we review the basic properties of Skyrmions, including
the winding number. In Sec. \ref{sec_coreR}, we study the
core of isolated Skyrmions in various phases
of the model. We find that for paramagnets and
ferromagnets, the core behavior is readily found. The core behavior
defines a measure of the core size, called $R$, which we find to be the
result of the competition of different physical length scales in
different regions of the phase diagram. We show that for the aligned
conical phase, the core behavior is undetermined because of the scale
invariance due to the spontaneous symmetry breaking. In
Sec. \ref{sec_tail}, we study the Skyrmion tails in various parts of
the phase diagram. For paramagnets and ferromagnets, we find that the
tail is exponentially decaying. The decay length of the
tail depends on various length scales in different regions of the
phase diagram. The Skyrmions in aligned conical
phase, however, are algebraically decaying for large distances due to the underlying Goldstone
modes in the system. In
Sec. \ref{sec_abrikosov}, we employ the Abrikosov flux
lattice to study the Skyrmion lattice. We find that the core size
depends on $q$ and the thermal correlation length of paramagnets. In Sec.
\ref{sec_exprL}, we match the core and tail of Skyrmions, with the
matching radius introduced as a new measure of the Skyrmion size. We argue
that this measure of Skyrmion size is the most physical. In
general, we find that the core size is of the order $q^{-1}$ near the
helimagnetic phase boundary, consistent
with the results of other previous studies, \cite{Bog95,BogHub94} but it depends on other length scales
in other parts of the phase diagram. We do not discuss
pinning, polarization, and the alignment effects mentioned in
a previous paper. \cite{HoKiSaBe10}

\section{Model and Length Scales} \label{sec_model}

\subsection{LGW functional}

Throughout this paper, we use the LGW model with DM Interaction \cite{HoKiSaBe10}
\begin{eqnarray}
\nonumber \mathcal{S}[\bM] &=& \int d^3 x \left[ \frac{r}{2} \bM^2 + \frac{a}{2}
  (\nabla \bM)^2 + \frac{c}{2} \bM \cdot \nabla \times \bM \right.\\
\label{expr_action} && + \left. \frac{u}{4} (\bM^2)^2 - \bH\cdot\bM \right] .
\end{eqnarray}
The terms with coefficients $r$, $a$ and $u$ are the usual
Landau-Ginzburg-Wilson (LGW) model. \cite{Ma76} The $c$-term is the
DM interaction \cite{Dzy58, Mor60}, and it exists in systems with
spin-orbit coupling \cite{BelKir10} and no inversion symmetry in the
unit cell of the solid. \cite{PluWal81}
A similar chiral
structure can be found in cholestoric liquid crystal. \cite{Lub72} $H$
is the external magnetic field. Its direction defines the $z$-axis
throughout the paper. The saddle-point equation corresponding to
Eq. (\ref{expr_action}) is \cite{FeHiSt10}
\begin{equation}
\label{saddlepteqn} r \bM - a \nabla^2 \bM + c \nabla \times \bM + u
M^2 \bM  -\bH = 0 ,
\end{equation}
We do not consider crystal-field effects in this paper. Generally
these terms align the helix in a certain direction for small magnetic
field. For large magnetic fields, the helix is in the direction of the field.

\subsection{Phases and Length Scales} \label{sec_phases}

\begin{figure}
\includegraphics[scale=0.25]{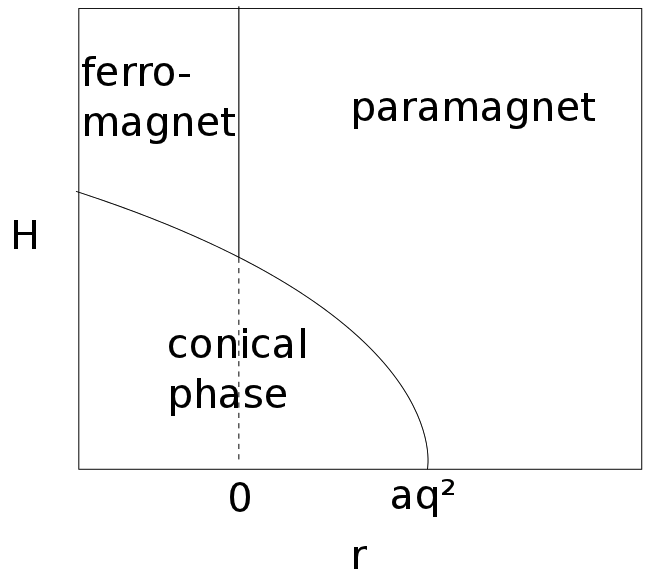}
\caption{\label{gr_phdiagLGWDM} Phase diagram predicted by the
  action in Eq. (\ref{expr_action}).}
\end{figure}

There are three topologically trivial phases that are associated with
the action in Eq. (\ref{expr_action}). These three phases are the paramagnet, the
ferromagnet and the aligned conical phase, as shown in
Fig. (\ref{gr_phdiagLGWDM}). The
aligned conical phase is stable only if $H \leq H_{c2}$ where $H_{c2}$
is defined below, see Eq. (\ref{expr_Hc2}).

\subsubsection{Paramagnet}

The paramagnet is stable or metastable only for $r>0$. Its analytic
form is given by Eq. (\ref{expr_MP}). In a magnetic field, the
magnetization is
\begin{equation}
\label{PM_smallH} m = \chi_p H,
\end{equation}
where $\chi_p = r^{-1}$ is the magnetic susceptibility. The
corresponding correlation
length is given by $\xi_p$, which diverges when $r$
approaches $0$. In LGW model, it is \cite{Ma76}
\begin{equation}
\label{def_xip} \xi_p = \sqrt{\frac{a}{r}},
\end{equation}
We define a length (basically the thermal and magnetic field dependent
transverse correlation length)
\begin{equation}
\label{def_barlH} \barlH = \sqrt{\frac{am}{H}} .
\end{equation}
For $r \gg \frac{am}{H}$, it is
\begin{equation}
\label{expr_barlH_PM} \barlH \approx \xi_p .
\end{equation}
But for $r \ll \frac{am}{H}$, $m \approx (u^{-1}
H)^{\frac{1}{3}}$. Hence the length becomes
\begin{equation}
\label{expr_barlH_TCP} \barlH \approx
\frac{a^{\frac{1}{2}}}{u^{\frac{1}{6}} H^{\frac{1}{3}}} ,
\end{equation}
which is related to the mean-field critical exponent $\delta=3$. \cite{Ma76}

\subsubsection{Ferromagnet}

The ferromagnet is stable or metastable only for $r<0$. Its analytic
form is given by Eq. (\ref{expr_MF}). Without external magnetic
field, the magnetization is given by
\begin{equation}
\label{expr_MFH0} M_F^{(0)} = \sqrt{\frac{|r|}{u}} .
\end{equation}
A magnetic field along the direction of the magnet gives
\begin{equation}
m \approx M_F^{(0)} + \chi_f H ,
\end{equation}
where $\chi_f = (2|r|)^{-1}$ is the longitudinal susceptibility. The
translational susceptibility goes like $H^{-1}$.

The longitudinal correlation length of the ferromagnet is given by
\begin{equation}
\label{def_xif} \xi_f = \sqrt{\frac{a}{2|r|}} ,
\end{equation}
which diverges for $r \rightarrow 0$.
The transverse correlation length is given by $\barlH$ in
Eq. (\ref{def_barlH}). It is infinite for zero magnetic
field, but in a field it is
\begin{equation}
\label{expr_barlH_FM} \barlH \approx \frac{|r|^{\frac{1}{4}}
  a^{\frac{1}{2}}}{u^{\frac{1}{4}} H^{\frac{1}{2}}} .
\end{equation}

\subsubsection{Aligned Conical Phase}

The aligned conical phase, or simply conical phase, contains a component of
helicity and another of homogeneous magnetization, given by \cite{HoKiSaBe10}
\begin{equation}
\label{def_MCP} \mathbf{M}_{CP} = m_{sp} (\cos qz \bunitx + \sin qz \bunity) + \ml \bunitz ,
\end{equation}
where $q$ is the pitch vector of the helix given by
\begin{equation}
\label{value_q} q = \frac{c}{2a} .
\end{equation}
It is proportional to the strength of DM interaction.
$\ml$ is the homogeneous magnetization induced by the magnetic field,
and is given by
\begin{equation}
\label{def_ml} \ml = \chi_h H ,
\end{equation}
where $\chi_h = (aq^2)^{-1}$ is the magnetic susceptibility of the homogeneous
magnetization. $m_{sp}$ is the helimagnetic amplitude given by
\begin{equation}
\label{def_msp} m_{sp} = \sqrt{m_H^2 - \ml^2} ,
\end{equation}
where
\begin{equation}
\label{def_mH} m_H = \sqrt{\frac{a q^2-r}{u}} .
\end{equation}

The aligned conical phase can be the thermodynamically stable phase as in
Fig. (\ref{gr_phdiagLGWDM}), if the system is in the region
\begin{equation}
\label{criterion_ACP} a q^2 - r \geq \frac{u H^2}{a^2 q^4} .
\end{equation}
This defines the critical field $H_{c2}$
\begin{equation}
\label{expr_Hc2} H_{c2} = a q^2 \sqrt{\frac{aq^2-r}{u}} .
\end{equation}

At zero field, the correlation length approaching the helimagnetic-paramagnetic
phase transition from the paramagnetic phase ($r>a q^2$) is given by
\begin{equation}
\label{def_xih} \xi_h = \sqrt{\frac{a}{r-aq^2}} =
\left(\frac{1}{\xi_p^2} - q^2 \right)^{-\frac{1}{2}} .
\end{equation}
If the phase transition is approaching from the helimagnetic phase ($r <
a q^2$), it is given by
\begin{equation}
\label{def_xihHM} \xi_h' = \sqrt{\frac{a}{aq^2-r}} =
\left(q^2 - \frac{1}{\xi_p^2} \right)^{-\frac{1}{2}} .
\end{equation}
At the transition point between the helimagnet and paramagnet at
$H=0$,
\begin{eqnarray*}
q\xi_p = 1.
\end{eqnarray*}

\section{Basic Properties of Skyrmions} \label{sec_basicsSkyr}

\subsection{Winding Number}

Skyrmions are two-dimensional objects, which are topologically non-trivial
because the winding number of a Skyrmionic configuration is
non-zero. Assume that
\begin{equation}
\bM(\bx) = m(\bx) \bn(\bx) ,
\end{equation}
where $m(\bx)$ and $\bn(\bx)$ denotes the magnitude and direction of
$\bM$. The winding number is defined as \cite{Raj87}
\begin{equation}
\label{def_W} W = \int dx \int dy \frac{1}{4\pi} \bn \cdot \left(\frac{\partial
    \bn}{\partial x} \times \frac{\partial \bn}{\partial y}\right)   .
\end{equation}
Upon continuous deformation of the configurations, the
winding number $W$ remains unchanged. Note that all phases in
Sec. \ref{sec_phases} have $W=0$, which means they are
all topologically trivial. 

\subsection{Description of Skyrmions}

\begin{figure}[htp]
\centering
\includegraphics[scale=.45]{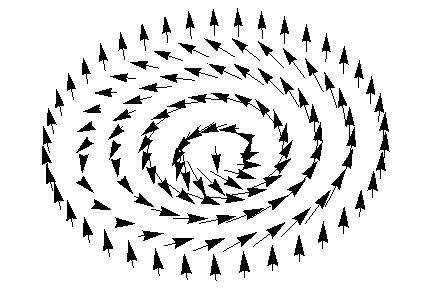}
\caption{The picture of an isolated azimuthal Skyrmion with
  $W=-1$.} \label{gr_singleskyrmion}
\end{figure}

To illustrate a topologically non-trivial solution, we write the
configuration in the form of \cite{GenPro93} (in cylindrical coordinates)
\begin{eqnarray}
\nonumber \bn(\bx) 
&=& \sin\theta(\bx) \cos\alpha(\bx) \bunitrho +
\sin\theta(\bx)\sin\alpha(\bx) \bunitphi + \cos\theta(\bx) \bunitz ,\\
\label{defbn} &&
\end{eqnarray}
so that the winding number
can be written as
\begin{widetext}
\begin{equation}
\label{W_ansatz} W = \frac{1}{4\pi} \int_0^{\infty} d\rho
\int_0^{2\pi} d\varphi \cdot \sin\theta(\bx) \left[ - \frac{\partial
    \theta(\bx)}{\partial \varphi} \frac{\partial
    \alpha(\bx)}{\partial \rho} + \left(1+\frac{\partial
      \alpha(\bx)}{\partial \varphi}\right) \frac{\partial
    \theta(\bx)}{\partial \rho}    \right] .
\end{equation}
\end{widetext}
For all the Skyrmions we review and present in this paper, $\theta(\rho=0) = \pi$ and
$\theta(\rho=\infty) = 0$, so that $W=-1$. In addition, we set
$\alpha=\frac{\pi}{2}$, meaning it is an azimuthal Skyrmion. We
exclude the consideration of a radial Skyrmion because the
presence of DM interaction forces the Skyrmions be azimuthal. The
winding number describes how the electron changes its
  spin when it passes through the core.  \cite{PflRos10} Such a
  Skyrmion is depicted in Fig. \ref{gr_singleskyrmion}.

Although Skyrmions are topological objects, linear response
ensures that the asymptotic
behaviors of the tails of Skyrmions are no different from the decay of
other kinds of perturbations. It is the characteristics of the core that
constitutes the topology. In understanding Skyrmions in various
ordered phases of the helimagnets, we study the core by differential
equations with boundary conditions that give a non-zero winding
number, and the tail through perturbation theory.

\section{Skyrmion Cores} \label{sec_coreR}

\subsection{Ferromagnet and Paramagnet} \label{sec_coreRFMPM}

We first explore the cases of paramagnets and ferromagnets. And as we have stated in Section
\ref{sec_basicsSkyr}, $\theta(\rho=0)=\pi$. Then we write, for
$\rho \rightarrow 0$,
\begin{eqnarray}
\label{perturbmcore} m(\bx) &=& m_{\infty} + \delta m(\rho) ,\\
\label{perturbthetacore} \theta(\bx) &=& \pi + \delta \theta(\rho) , 
\end{eqnarray}
where all behavior depends only on the radial distance $\rho$, and
where $m_{\infty} = M_P$ (for paramagnet, $r>0$, as in Eq. (\ref{expr_MP})) or $M_F$ (for
ferromagnet, $r<0$, as in Eq. (\ref{expr_MF})). Putting Eq.
(\ref{perturbmcore}) and Eq. (\ref{perturbthetacore}) in the saddle-point
equations (\ref{saddlepteqnPMFM1}) and (\ref{saddlepteqnPMFM2}),
ignoring all higher order terms, we can find that for $\rho
\rightarrow 0$, $\delta \theta = \Theta_c \rho$ and $\delta m = M_c
\rho^2$, with
\begin{eqnarray}
\label{expr_Mc} M_c &=& \frac{m_{\infty}}{2
  \barlH^2 (1 +
  4q^2 \barlH^2)} ,\\
\label{expr_Thetac} \Theta_c &=& - \frac{2q}{1+4q^2 \barlH^2} ,
\end{eqnarray}
where $\barlH$ is the transverse correlation length for paramagnets
or ferromagnets in Eq. (\ref{expr_barlH_PM}), Eq. (\ref{expr_barlH_TCP})
or Eq. (\ref{expr_barlH_FM}).
Note that this calculation breaks down if there is no DM interaction, or the
Skyrmion is in an ordered phase.

The size of the core can be estimated with this solution as $R$, where $\theta =
\pi \left(1-\frac{\rho}{R}\right)$. Let us consider different cases in ferromagnets and paramagnets, and
the relation of Skyrmion in paramagnets in some situations to the
Skyrmion lattice.

\subsubsection{Ferromagnet}

For a ferromagnet with non-zero magnetic field, the core size $R$
is given by the following cases (with the phase stable or metastable):
\begin{enumerate}
\item $\barlH \ll q^{-1}$: $R \approx \frac{\pi}{2q}$.
\item $q^{-1} \ll \barlH$: $R \approx
  2\pi q \barlH^2$.
\end{enumerate}
The first case refers to the region where the magnetic field is much
larger than the critical field $H_{c2}$. The core size is proportional
to $q^{-1}$. The second case refers to the region closer to $H =
H_{c2}$, where the magnetic length $\barlH$ plays a role.

\subsubsection{Paramagnet}

For a paramagnet, the core size $R$
is given by (in which the phase can be stable or
metastable):
\begin{enumerate}
\item $\xi_p \ll q^{-1}$: $R \approx \frac{\pi}{2q}$.
\item $\xi_p \approx q^{-1}$ and along $H \approx H_{c2}$: $R \approx \frac{5\pi}{2q}$.
\end{enumerate}
The magnetic field does not play a role in the core behavior
for paramagnets. It is because close to $r \approx 0$, the paramagnet
appears for $H > H_{c2}$ and $\barlH$ is less significant than the
contribution of $q^{-1}$. For $r > \frac{am}{H}$, $\barlH \approx \xi_p$. 

\subsection{Aligned Conical Phase}

The core behavior for aligned conical phase is different from
paramagnets and ferromagnets in Sec. \ref{sec_coreRFMPM} because of
the Goldstone mode in this phase. To understand
it, we first study ferromagnet without external magnetic field and DM
interaction. This is for illustrative purpose, because it breaks a continuous
symmetry just like the conical phase does.

\subsubsection{Ferromagnet without external magnetic field and DM
  interaction}  \label{sec_coreFMH0q0}

With $H=0$ and $q=0$, the ferromagnet (with $r<0$) is an
ordered phase that breaks the continuous rotational symmetry of the
action in Eq. (\ref{expr_action}). The differential equation for
$\delta \theta$ in Eq.
(\ref{perturbthetacore}) as in Eq. (\ref{saddlepteqnPMFM2}) becomes
\begin{eqnarray}
\label{dtheta_FMcore} \frac{d^2 \delta\theta}{d\rho^2} + \frac{1}{\rho}
\frac{d\delta\theta}{d\rho} - \frac{1}{\rho^2} \delta\theta = 0.
\end{eqnarray}
The differential equation is due to the gradient term in the
action. Then
\begin{equation}
\label{expr_deltatheta_coreDM} \delta \theta = A \rho,
\end{equation}
with an arbitrary coefficients $A$. In fact, the approximation in Eq. (\ref{SkyrNLSM_approx})
from the exact Skyrmion solution to the non-linear $\sigma$ model 
near the core gives the same. Therefore, we cannot determine the core size. To determine the core size
just from the core behavior, external
perturbations that provide extra length scales are needed. For example, the presence of
DM interaction fixes $l$ in Eq. (\ref{SkyrNLSM_approx}) to be $(4q)^{-1}$.

\subsubsection{Aligned Conical Phase} \label{sec_coreACP}

The aligned conical phase breaks the continuous translational symmetry
of the action in Eq. (\ref{expr_action}). \cite{HoKiSaBe10} Because of the
helical nature of this phase, we expect $m(\bx)$, $\theta(\bx)$ and
$\alpha(\bx)$ depend on $\xi$ (defined in Eq. (\ref{def_xi})) in
addition to the radial distance $\rho$. After analyzing
Eqs. (\ref{saddlepteqnHM1}), (\ref{saddlepteqnHM2}) and
(\ref{saddlepteqnHM3}), the dominant variation near the core is $\rho \sin
\xi$. However, the coefficients is undetermined for the same reason as
Sec. \ref{sec_coreFMH0q0}.  The perturbation for
conical phase can be written as Eq. (\ref{ACPeigenmode}). The
differential equations for the fluctuations for conical phase
can be written as a Laplace equation as in Eq. (\ref{PDEACPGS}), although both gradient terms and
curl term (due to DM interaction) in the action in
Eq. (\ref{expr_action}) are important for the aligned conical
phase.

\section{Skyrmion Tails} \label{sec_tail}

The tail behavior of the Skyrmions is closely related to the
correlations in the bulk of the phase. \cite{For89} This can be shown by linear response theory (see
Appendix \ref{app_LRTSPE}).  In the following, we show by each case
that the presence of spontaneous symmetry breaking gives the Skyrmion an
algebraic tail, but an exponentially decaying tail otherwise. The decay length
of the tail is in general not the same as the core size. 

\subsection{Paramagnet and Ferromagnet} \label{sec_tailPMFM}

We study the system using perturbation techniques. As we have stated in Section
\ref{sec_basicsSkyr}, $\theta(\rho=\infty)=0$. Then we write, for
$\rho \rightarrow \infty$,
\begin{eqnarray}
\label{perturbmfar} m(\bx) &=& m_{\infty} + \delta m(\rho) ,\\
\label{perturbthetafar} \theta(\bx) &=& 0 + \delta \theta(\rho) , 
\end{eqnarray}
where all behavior depends only on the radial distance $\rho$, and
$m_{\infty} = M_P$ (for paramagnet, $r>0$, as in Eq. (\ref{expr_MP})) or
$M_F$ (for ferromagnet, $r<0$, as in Eq. (\ref{expr_MF})) as in
Sec. \ref{sec_coreR}. We expect that the tail is exponential. Hence,
we assume $\delta m = \tilde{M} e^{-K \rho}$ and
$\delta \theta = \Theta e^{-K \rho}$. 
One of the solutions for $K$, called $K_-$, is used to define the
length of the tail
\begin{equation}
\label{def_lT} l_T = \frac{1}{|K_-|},
\end{equation}
which is another measure of the size of Skyrmions. In some cases,
$K_-$ has an imaginary part, which indicates the tails are oscillating
in addition to the exponential decay, but we will omit oscillations
below despite its existence in some cases.

\subsubsection{Ferromagnet}

For ferromagnet, the lengths of the Skyrmion tails are:
\begin{enumerate}
\item $\barlH \ll q^{-1} \ll \xi_f$: $l_T \approx \barlH$.
\item $\barlH \ll \xi_f \ll q^{-1}$ and $\xi_f \ll \barlH \ll q^{-1}$:
  $l_T \approx \barlH$.
\end{enumerate}
For both cases, the tails has a length $l_T \sim H^{-\frac{1}{2}}$ as
in Eq. (\ref{expr_barlH_FM}). We expect this
because the transverse fluctuations of ferromagnet have the spectrum
$\omega(\bk) = k^2 + H$. \cite{HoKiSaBe10}

\subsubsection{Paramagnet}

For paramagnet, the lengths of the tails are:
\begin{enumerate}
\item $\barlH \ll q^{-1} \ll \xi_p$: $l_T \approx \barlH$.
\item $\xi_p \ll q^{-1} \ll \barlH$: $l_T \approx \xi_p$.
\item $\xi_p \lesssim q^{-1}$ and small $H$: $l_T \approx \xi_h$.
\item Along $H \approx H_{c2}$: $l_T \approx \frac {\sqrt{2}}{\kappa_-}$,
\end{enumerate}
where for the fourth case,
\begin{equation}
\label{def_kappa} \kappa_{\pm} =  \sqrt{q \sqrt{q^2+\frac{2}{\xi_h'^2}} \pm \left(q^2 -
      \frac{1}{\xi_h'^2}\right)} .
\end{equation}

The first case corresponds to the the paramagnet with large magnetic
field and the boundary with ferromagnet, where $\barlH$ is given by
Eq. (\ref{expr_barlH_PM}). The second case refers to the paramagnet
far away from the transition points, making $l_T \approx \xi_p$. The
third case refers to the paramagnet very close to the
helimagnetic transition point, making $l_T \approx \xi_h$.

The fourth case refers to the paramagnet along $H \approx H_{c2}$. The
Skyrmion lattice is formed along part of this critical line.

\subsection{Aligned Conical Phase}

By Goldstone theorem and linear response, any perturbation in the conical phase shows
long distance algebraic decay. Therefore, a
Skyrmion in this phase shows a long tail. For
illustrative purpose, we study the ferromagnet without magnetic field
and DM interaction which breaks the rotational symmetry.

\subsubsection{Ferromagnet without external magnetic field and DM
  interaction}  \label{sec_tailFMH0q0}

The spectrum of the 
Goldstone modes in ferromagnet in $H=0$ and $q=0$ is $\omega(\bk) = k^2$, \cite{Ma76}
and the modes are readily diagonalized as $\delta m_x$ and
$\delta m_y$. As a result, they behaves $|\mathbf{r}|^{-1}$, as
illustrated in Appendix \ref{app_LRTSPE}. We expect the Skyrmion tail
to behave in the same way.

We still employ the perturbation schemes in Eq. (\ref{perturbmfar})
and Eq. (\ref{perturbthetafar}) for this ferromagnet with $H=0$ and
$q=0$. The differential equations kept to the relevant order is given
as
\begin{eqnarray}
\nonumber \frac{\partial^2 \delta
      m}{\partial \rho^2} + \frac{1}{\rho} \frac{\partial \delta
      m}{\partial \rho} - M_F^{(0)} \left(\frac{\partial \delta
      \theta}{\partial \rho}\right)^2 - \frac{M_F^{(0)}}{\rho^2} (\delta
  \theta)^2 &=& \frac{\delta m}{\xi_f^2}  ,\\
\label{eqn_powerlaw_FM1} && \\
\label{eqn_powerlaw_FM2} \frac{\partial^2 \delta\theta}{\partial \rho} + \frac{1}{\rho}
\frac{\partial \delta\theta}{\partial \rho} - \frac{1}{\rho^2}
\delta\theta &=& 0 .
\end{eqnarray}
From an analysis of the differential equations
(\ref{saddlepteqnPMFM1}) and (\ref{saddlepteqnPMFM1}), we get
\begin{eqnarray}
\label{expr_deltatheta_farcoreFM} \delta \theta &=& \frac{\Theta_f}{\rho} ,\\
\label{expr_deltam_farcoreFM} \delta m &=& - 2 \xi_f^2 M_F^{(0)} \frac{\Theta_f^2}{\rho^4} .
\end{eqnarray}
From Eq. (\ref{eqn_powerlaw_FM2}), we know that the coefficients
$\Theta_f$ is arbitary, as in core solution the
coefficients in Eq. (\ref{expr_deltatheta_coreDM}) in Sec. \ref{sec_coreFMH0q0} is undetermined as
well. It is the result of the equation for the fluctuations given by a
Laplace equation for a symmetry-breaking ferromagnet. $\delta m(\rho)$ and $\delta \theta(\rho)$ can
be expressed in the scaling form $f\left(\frac{\xi_f^2}{L^2}
  ,\frac{\rho}{L}\right)$, for some length scale $L$.
Moreover, the approximation in Eq. (\ref{SkyrNLSM_approx})
far from the Skyrmion core in the non-linear $\sigma$ model has the
same behavior. Similarly, as in the Skyrmion core, the coefficients can be fixed by additional
interactions that carry other length scales as discussed in Sec. \ref{sec_coreFMH0q0}.

\subsubsection{Aligned Conical Phase} \label{sec_tailACP}

While the ferromagnet has the readily diagonalized
Goldstone modes with spectrum $\omega(\bk) = k^2$, the aligned conical phase
has the Goldstone modes given by \cite{HoKiSaBe10}
\begin{equation}
\label{Goldstone_ACP} \omega(\bk) \approx k_z^2 +
\frac{\xi_h'^{-2}}{q^2+\xi_h'^{-2}} \left(\frac{H}{H_{c2}}\right)^2 \bkper^2 +
 \left[1-\frac{4}{q^2 \xi_h'^2} \left(\frac{H}{H_{c2}}\right)^2 \right] \frac{\bkper^4}{2q^2} ,
\end{equation}
for small magnetic field ($\ml^2 \ll m_H^2$), similar to that of the
cholesteric liquid crystal. \cite{Lub72} For zero magnetic field
makes, the second term vanishes. \cite{BeKiRo06a}
For the parametrization of
fluctuations about the conical phase (\ref{def_MCP}) can be written as \cite{HoKiSaBe10}
\begin{eqnarray}
\nonumber \bM(\bx) &=& m_{sp} \left[\cos(qz+\varphi_0(\bx)) \bunitx +
\sin(qz+\varphi_0(\bx)) \bunity \right.\\
\nonumber && \left. + (\varphi_+(\bx) \cos qz +
\varphi_-(\bx) \sin qz) \bunitz \right] \\
\label{perturbM} && + \ml [\pi_1(\bx) \bunitx + \pi_2(\bx) \bunity + \bunitz] ,
\end{eqnarray}
to the first order of all parameters. The Goldstone mode that
corresponds to (\ref{Goldstone_ACP}) is given by the
``diagonalized'' form as
\begin{eqnarray}
\nonumber && \delta m_{G} (\bx) \\
\nonumber  &\approx& \phi_0(\bx)  - \frac{q^2+\xi_h'^{-2} \left[1- \left(\frac{H}{H_{c2}}\right)^2\right]}{q (q^2+\xi_h'^{-2})} \frac{\partial
  \varphi_+(\bx)}{\partial y} \\
\nonumber && + \frac{q^2+\xi_h'^{-2} \left[1 - \left(\frac{H}{H_{c2}}\right)^2\right]}{q (q^2+\xi_h'^{-2})} \frac{\partial
  \varphi_-(\bx)}{\partial x} \\
\label{ACPeigenmode} && - \frac{\xi_h'^{-2} \left[1- \left(\frac{H}{H_{c2}}\right)^2\right]}{q (q^2+\xi_h'^{-2})} \frac{\partial
  \pi_1(\bx)}{\partial x}  + \frac{\xi_h'^{-2} \left[1- \left(\frac{H}{H_{c2}}\right)^2\right]}{q (q^2+\xi_h'^{-2})} \frac{\partial
  \pi_2(\bx)}{\partial y} ,
\end{eqnarray}
which satisfies the partial differential equation
\begin{equation}
\label{PDEACPGS} \left[\frac{\partial^2}{\partial z^2}
+\frac{\xi_h'^{-2}}{q^2+\xi_h'^{-2}} \left(\frac{H}{H_{c2}}\right)^2
\nabla_{\bot}^2\right] \delta m_G(\bx) \approx 0 ,
\end{equation}
for $H \neq 0$. \footnote{We do not consider the case for $H=0$ because, while
it does have a partial differential equation for $\delta m_G$, it
is thermodynamically unstable as $\int d^3\bx e^{i \bk\cdot\bx} \left(k_z^2 + \frac{\bkper^4}{2 q^2}\right)^{-1} =
\sqrt{2}\pi q \int d^2\bkper \frac{1}{\bkper^2} e^{i \bkper \cdot
  \bxper} e^{- \frac{\bkper^2}{\sqrt{2} q} z}$, which is essentially equivalent to the Goldstone modes in $d=2$, which
leads to logarithmic divergence, according to Mermin-Wagner
theorem. \cite{Col73,Hoh67,MerWag66}.}
Eq. (\ref{PDEACPGS}) has a solution
\begin{equation}
\label{formflucACP} \delta m_G(\bx) =  \frac{A}{\rho} \sin(\varphi +
B) ,
\end{equation}
with undetermined coefficients $A$ and $B$. (Note that the Skyrmion core in conical phase has
arbitrary size $R$ and goes like $\rho \sin(\varphi+B)$ for the same argument.) Therefore,
similar to ferromagnets, the Skyrmion tail in the conical phase has a
power law form. A detailed analysis of the saddle-point equations in
Appendix \ref{app_saddlepteqnHM} shows that Skyrmion tail in the
conical phase goes like $\rho^{-1} \sin\xi$, where $\xi$ is defined in Eq.  (\ref{def_xi}),
with arbitrary coefficients.

\section{Skyrmion Lattice} \label{sec_abrikosov}

In this section, we study a lattice of Skyrmion in helimagnets. It is
convenient to use $CP^1$ representation because Skyrmions are like
vortices in superconductors and the technique of Abrikosov lattice
of vortices can be used. The details of this representation can be found
in Appendix \ref{app_CP1}. The saddle-point equations in this
representation are Eq. (\ref{saddleeqn1CP1}) and Eq. (\ref{saddleeqn2CP1}).

\begin{figure}
\includegraphics[scale=0.5]{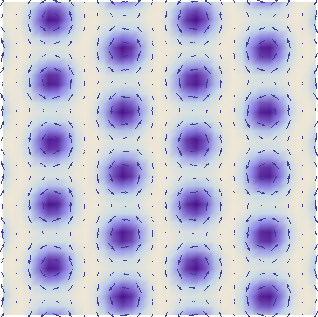}
\caption{\label{gr_abrlat2plane} Hexagonal Skyrmion lattice described by
  (\ref{expr_abrlat}), for $h \sim q^2$, where the vectors denote the
  projection of the spin on the plane, and the color denotes $n_z =
  z_i^* \sigma_z^{ij} z_j$ where deep blue denotes spin reversed from
  the magnetic field.}
\end{figure}

In Eq. (\ref{saddleeqn2CP1}), the gauge $\mathbf{A}$ depends on $\bz$
as in Eq. (\ref{def_gauge}),
leading to the non-linearity. To fix the gauge, set
$\mathbf{A} = -h x \bunity$. The dimension of $h$ is that of the
reciprocal of area, and
it will be shown later that it is related to the area of a
Skyrmion core. Because of the periodic nature of the
lattice, the term $-i q m n_{\alpha} \partial_{\alpha} z_i \rightarrow
-i q m \langle n_{\alpha} \rangle \partial_{\alpha} z_i$ is
ignored. This can be justified by the final solution. The
term $-\frac{i}{2} q m z_i \partial_{\alpha}
n_{\alpha}$ is also zero because the Skyrmions are azimuthal. Moreover, instead of keeping strictly
$\bz^{\dag} \bz = 1$, we relax the condition to $\langle \bz^{\dag}
\bz \rangle = 1$ where the average is
over one lattice. Following Abrikosov, \cite{Abr57} one
part of the solution is given by \cite{HZYPN10}
\begin{equation}
\label{expr_singlesite} \bz = \sqrt{\frac{\frac{h}{3^{\frac{1}{4}} \pi}}{1+|d_0|^2}} e^{iky} e^{-\frac{h}{2} \left(x+\frac{k}{h}\right)^2}\left[ \begin{array}{c}
    1 \\
d_0 \sqrt{2h} \left(x+\frac{k}{h}\right) \end{array} \right] ,
\end{equation}
where the prefactor is for normalization, and 
\begin{eqnarray*}
d_0 = - \frac{iq}{\sqrt{2h}} \frac{1}{\frac{1}{2} + \frac{H}{4ahm} +
\sqrt{\frac{1}{4} + \frac{q^2}{2h} + \frac{H}{4ahm} +
  \left(\frac{H}{4ahm}\right)^2} } .
\end{eqnarray*}
Let $l_x$ and $l_y$ be the distances between
cores along the  $x$ and $y$ axes respectively, where $l_x l_y =
\frac{2\pi}{h}$. Then $\bz$ can be seen as the superposition of the
above solution with different values of $k$ where $k_j = \frac{2\pi
  j}{l_y}$, then 
\begin{equation}
\label{expr_abrlat} \bz = \sqrt{\frac{\frac{h}{3^{\frac{1}{4}} \pi}}{1+|d_0|^2}} \sum_{j=-\infty}^{\infty} c_j
e^{i \frac{2\pi j}{l_y} y}  e^{-\frac{h}{2} \left(x+j l_x \right)^2} \left[
\begin{array}{c}
1 \\
d_0 \sqrt{2h} \left(x+j l_x \right)
\end{array}
 \right] .
\end{equation}
For a triangular lattice, $c_j = c_{j+2}$ \cite{ann04}. Choose $c_j$
to be $\frac{1}{\sqrt{2}}$ and $\frac{i}{\sqrt{2}}$ for even and odd
$j$'s. And $l_y = \frac{\sqrt{3}}{2} l_x$. Such configuration is plotted as shown in
Fig. \ref{gr_abrlat2plane} for $h \sim q^2$, which denotes a Skyrmion
lattice. On the other hand, a graph with spin projected on the basal
plane ($x$-$y$ plane) and a density plot of $n_z =
  z_i^* \sigma_z^{ij} z_j$ is plotted in Fig. \ref{gr_abrlat2plane}. 

To know the magnetization and the core size, we have to put Eq.
(\ref{expr_abrlat}) back to Eq. (\ref{expr_action_CP1}) and determine them
by variational method. Since the lattice is periodic, it is valid
and convenient to consider the solution of a single site in Eq.
(\ref{expr_singlesite}). Define
\begin{eqnarray*}
\tilde{h} &=& \frac{h}{q^2} ,\\
\cal{D} &=& \left[ \frac{1}{2} + \frac{H}{4 a q^2 m \tilde{h}}
\right. \\
&& \left. + \sqrt{
    \frac{1}{4} + \frac{1}{2 \tilde{h}} + \frac{H}{4 a q^2 m
      \tilde{h}} + \left(\frac{H}{4 a q^2 m
      \tilde{h}}\right)^2 }\right]^{-1} ,
\end{eqnarray*}
the free energy per unit volume of one Skyrmion in the lattice is
given by
\begin{equation}
\label{energy_abrlatsingle} \frac{\cal{F}}{V} = \frac{r}{2} m^2 + \frac{u}{4} m^4 + a q^2
  m^2 \frac{ \tilde{h} \left( 1 +
    \frac{3 {\cal{D}}^2}{2\tilde{h}} \right) - 2 \cal{D} }{1+\frac{{\cal{D}}^2}{2\tilde{h}}} - H m
\frac{1-\frac{{\cal{D}}^2}{2 \tilde{h}}}{1+\frac{{\cal{D}}^2}{2\tilde{h}}} .
\end{equation}
Then we evaluate magnetization $m$ and the reciprocal of core area $h$
by minimizing the free energy. There exists no analytic closed form
solution for $m$ and $h$, but we do it by qualitative
analysis. We expect that $m$ is of the same order of magnitude of
$\ml$ in the aligned conical phase or the paramagnet, and
$\frac{h}{q^2} < 1$. Expanding Eq. (\ref{energy_abrlatsingle}) for small
$\frac{h}{q^2}$, we get
\begin{eqnarray*}
\frac{\cal{F}}{V} \approx \frac{r}{2} m^2 + \frac{u}{4} m^4 - H m + a
q^2 m^2 \left(\frac{h}{q^2}\right) - \frac{4 a^3 q^6 m^4}{H^2}
\left(\frac{h}{q^2}\right)^2 .
\end{eqnarray*}
Minimizing it with respect to $h$ and $m$, we get
\begin{eqnarray*}
\frac{h}{q^2} \approx \frac{H^2}{8 a^2 q^4 m^2} ,\\
r m + u m^3 - H \approx 0 .
\end{eqnarray*}
The second equation indicates that the magnetization is approximately
equal to the paramagnet or ferromagnet. For small magnetic field and
in the paramagnetic phase ($r>0$), using Eq. (\ref{expr_MP}), we get
\begin{equation}
h \approx \frac{1}{8 q^2 \xi_p^4} .
\end{equation}
Therefore, the core-to-core distance goes like $q
\xi_p^2$. Near the helimagnetic phase boundary $\xi_p q \approx 1$, and
in this region the core-to-core distance goes like $q^{-1}$.

The experimentally
observed $A$ phase \cite{TPSF97} that has been
identified as hexagonal lattice of
Skyrmions is observed along the helimagnet/paramagnet phase boundary,
\cite{MBJPRNGB09,YOKPHMNT10}. Our result shows that the core size $\sim q \xi_p^2$,
which is $\sim q^{-1}$ for $q\xi_p \approx 1$, agreeing
with previous theoretical \cite{BogHub94,RoBoPf06,RoLeBo11,MBJPRNGB09,HZYPN10} and
experimental studies. \cite{YOKPHMNT10,TPSF97} However, we also
predict that the size increases
away from the phase boundary, provided 
the Skyrmion lattice is still the thermodynamic ground state when the
correlation length $\xi_p$ gets larger.

The fluctuations due to the Skyrmion lattice has the same form of that
of columnar phase in liquid crystal, \cite{KirBel10} and have the form
$\omega(\bk) = \bkper^2 + c_1 k_z^2 + c_2 k_z^4$. \cite{HoKiSaBe10}

\section{Core Size As the Matching Distance Between Core Behavior and
  Skyrmion Tail} \label{sec_exprL}

The definition of $R$ in Sec. \ref{sec_coreRFMPM} 
deals only with the core behavior. Here we introduce a new distance $L$,
which is defined as tbe distance where the core behavior and
tail of the Skyrmion meet. By finding
$L$, we consider both the core and the tail of the Skyrmion. In some
cases, $R$ and $L$ are not too different in
terms of order of magnitudes, but their difference becomes greater
when the magnetic field becomes large.

In the following cases, we match the core behavior and the tail at a
point $L$, and then we solve for $L$.

\subsection{Paramagnet and Ferromagnet}

For paramagnet and ferromagnet, by matching the core behavior in
Sec. \ref{sec_coreR} and the tail in
Sec. \ref{sec_tail}, we solve for the matching point $L$ tabulated in
Table \ref{tbl_PMFMskyrsol}. We verify that for all cases in Table
\ref{tbl_PMFMskyrsol} have winding
number $W=-1$, by putting the solutions of 
$\theta(\bx)$ back to Eq. (\ref{W_ansatz}).

\begin{table*}
\caption{\label{tbl_PMFMskyrsol} Isolated Skyrmions in
  ferromagnets and paramagnets in different regions
of the phase diagram in Fig. \ref{gr_phdiagLGWDM}. (PM = paramagnet, FM = ferromagnet)}
\begin{ruledtabular}
\begin{tabular}{|c|c|c|c|c|} 
\hline
 PM/FM & Region in Phase Diagram & $l_T$ & $R$ & $L$ \\
\hline
\hline
FM & $\xi_f \ll q^{-1} \ll \barlH$ & $\barlH$ & $2 \pi q \barlH^2$ & $\sqrt{32 \pi q^3 \barlH^3} \xi_f$ \\
\hline
FM & $\xi_f \ll \barlH \ll q^{-1}$ &
$\barlH$ & $\frac{\pi}{2q}$ & $\sqrt{4\pi q \barlH} \xi_f$ \\
\hline
FM & $\barlH \ll \xi_f \ll q^{-1}$ & $\barlH$ & $\frac{\pi}{2q}$ & $\sqrt{2\pi q \barlH} \barlH$ \\
\hline
FM & $\barlH \ll q^{-1} \ll \xi_f$ & $\barlH$ & $\frac{\pi}{2q}$ & $\sqrt{2 \pi q \barlH} \barlH$ \\
\hline
PM & $\xi_p \ll q^{-1} \ll \barlH$ & $\xi_p$ & $\frac{\pi}{2q}$ & $\frac{[2\pi (q\xi_p)^2]^{\frac{1}{3}}}{q}$ \\
\hline
PM & $q^{-1} \ll \barlH \ll \xi_p$ & $q^{-2} \barlH^{-1}$ & $2\pi q
\barlH^2$ & $2\pi q \barlH^2$ \\
\hline
PM & $\barlH \ll q^{-1} \ll \xi_p$ & $\barlH$ & $\frac{\pi}{2q}$ & $\sqrt{2\pi q \barlH} \barlH$ \\
\hline
PM & $\xi_p \lesssim q^{-1}$ & $\xi_h$ & $\frac{5\pi}{2q}$ & $1.48 q^{-1}$ \\
\hline
PM & Along $H \approx H_{c2}$ & $\frac {\sqrt{2}}{\kappa_-}$ & $\frac{5\pi}{2q}$ &
$\frac{\pi}{\sqrt{2} \kappa_+}$ \\
\hline
\end{tabular}
\end{ruledtabular}
\end{table*}

\begin{figure}[htp]
\centering
\includegraphics[scale=.25]{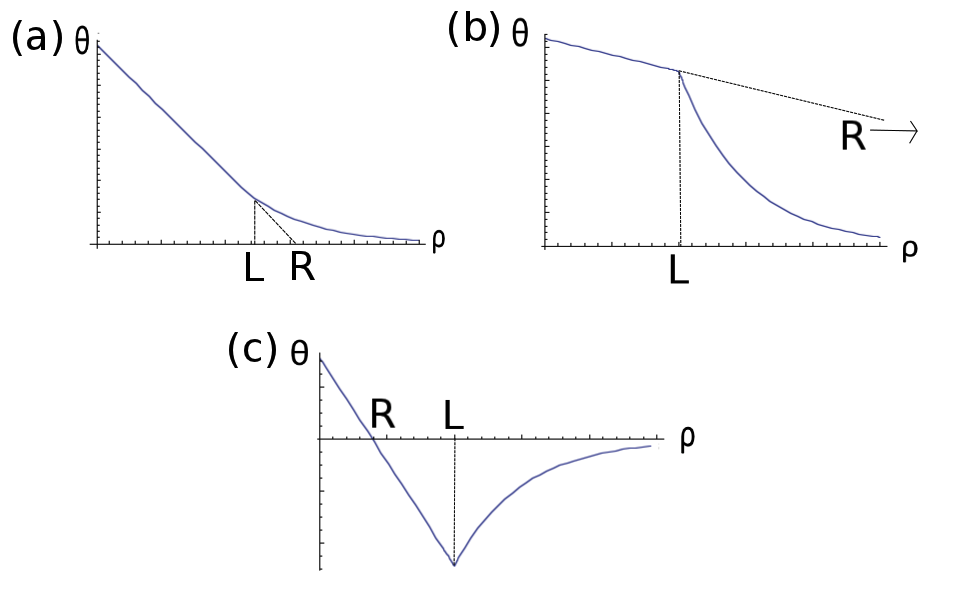}
\caption{Plots for $\theta(\rho)$ for Skyrmions in paramagnets and
  ferromagnets, and the meanings of $R$ and $L$ as the sizes of the Skyrmion
  core. (a) $R \approx L$. Both are equally good measures for the core
size. (b) $R \gg L$. $L$ is a better measure for the core size. (c) $R
\ll L$. $R$ is a better measure for the core size.} \label{gr_smRandL}
\end{figure}

From Table \ref{tbl_PMFMskyrsol}, we can see that $R$'s are mostly
of the order of magnitude of $q^{-1}$, and $l_T$'s are mostly the
thermal correlation lengths. However, $L$ shows much more complicated
dependence on the various length scales. $l_T$ is generally not a good
measure of a Skyrmion size because the correlation length is related
to the thermodynamic phase of the bulk, the size of an additional object. Both
$R$ and $L$ is of the order of magnitude of $q^{-1}$ near the
helimagnetic transition point at $H \approx 0$, indicating that 
the Skyrmion size is of $q^{-1}$ in this region. Far from this point,
$L$ and $R$ differs in orders of magnitude. In general, $L$ is better to
characterize the size of Skyrmions because it takes into
account both the core and the tail. Whether $L$ or $R$ is a better measure
depends on the situations, as listed below:
\begin{enumerate}
\item If $L$ and $R$ are of the same order of magnitude ($\sim K^{-1}$
  or oscillating) as in Fig. \ref{gr_smRandL} (a), they are
  equally good. Examples: paramagnets in 
  $q^{-1} \ll \barlH \ll \xi_p$ and $\xi_p \lesssim q^{-1} $.
\item If $R \gg L$ as in Fig. \ref{gr_smRandL} (b), $L$ is a better measure because $L$ depicts where
  the tail starts and the slope of the core behavior was
  underestimated. Examples: ferromagnets in $\xi_f \ll q^{-1} \ll
  \barlH$, $\xi_f \ll \barlH \ll q^{-1}$, $\barlH \ll \xi_f \ll q^{-1}$ and $\barlH
  \ll q^{-1} \ll \xi_f$, and paramagnets in $\barlH \ll q^{-1} \ll
  \xi_p$ and $\xi_p \ll q^{-1} \ll \barlH$.
\item If $L \gg R$ as in Fig. \ref{gr_smRandL} (c), $R$ is a better measure. The matching method is
  not working so well because at $\rho = L$, $\theta(\rho)$ becomes
  negative. However, There are no such examples in all cases
  considered in Table \ref{tbl_PMFMskyrsol}.
\end{enumerate}
Therefore, for our purpose, $L$ is a better characterization of the Skyrmion size in
general.

For pure ferromagnets ($q=0$ and $H=0$) and aligned conical phase,
matching does not fix the size of the Skyrmion due to the same reason
stated in Sec. \ref{sec_tailFMH0q0}. For pure ferromagnets, matching
the solutions in Eq. (\ref{SkyrNLSM_approx}) for small and large
$\rho$ does not give $L$.

\subsection{Aligned Conical Phase} \label{sec_matchingACP}

Because of the technical complexities of aligned conical phase, we do
not match the core and tail solutions as we did for other
magnets. However, we assume a form of Syrmionic solution here and
estimate the range of the core size. We describe
an isolated Skyrmion in terms of
\begin{eqnarray}
\nonumber \bM = [1-\eta_1(\rho,\xi)] m_{sp}  (\cos qz \bunitx + \sin
qz \bunity) \\ 
\label{bM_ACP_Sky} + \ml \bunitn_{sk} - \eta_3(\rho,\xi) \ml \bunitz ,
\end{eqnarray}
where $\bunitn_{sk}$ is the same Skyrmion in
(Eqs. \ref{sigmaSkyx}-\ref{sigmaSkyz}). The first and third
term are with variational parameters $\eta_1$ and $\eta_3$ which does
not alter the winding number. The second term with $\bunitn_{sk}$ keeps
the winding number to be $-1$. Because Eq. (\ref{bM_ACP_Sky})
already captures the long-range behavior as $\bunitn_{sk}\cdot (-\sin qz
\bunitx + \cos qz \bunity) \sim \frac{1}{\rho} \sin\xi$ far from the
core ($\rho \rightarrow \infty$), there is no
term in the direction of $(-\sin qz \bunitx + \cos qz \bunity)$ in Eq.
(\ref{bM_ACP_Sky}).

We then solve for $\eta_1$ and $\eta_3$ by putting Eq. (\ref{bM_ACP_Sky})
to the saddle-point equation (\ref{saddlepteqn}) for regions far from
the core ($\rho \rightarrow \infty$) and near the core ($\rho \approx
0$). Far from the core ($\rho \rightarrow \infty$), we find that
\begin{equation}
\label{eta1far} \eta_1 \approx \frac{(3aq^2-2r) \ml}{2 u m_{sp}^3}
\left(\frac{2l}{\rho}\sin\xi\right) ,
\end{equation}
\begin{equation}
\label{eta3far} \eta_3 \approx - \frac{\ml}{m_{sp}}
\left(\frac{2l}{\rho} \sin\xi\right) ,
\end{equation}
where $l$ is an undetermined parameter that we estimate below.
At the core ($\rho \approx 0$), with linearization of the parameters
in the saddle-point equation (\ref{saddlepteqn}), we find that
\begin{eqnarray}
\label{eta1near} \eta_1 \approx - E_1 \left(\frac{2\rho}{l}
  \sin\xi\right),\\
\label{eta3near} \eta_3 \approx 1 - B_3 \left(\frac{\rho}{l}\right)^2 - E_3
\left(\frac{2\rho}{l} \sin\xi \right),
\end{eqnarray}
where (in terms of bare parameters in the action)
\begin{eqnarray}
\label{exprE1} E_1 &=& \frac{\left[r (1-lq) + \frac{4a}{l^2} (2-q^2
    l^2)\right] \ml}{r-\frac{2aq}{l} + u (3 m_{sp}^2+4 \ml^2)} \\
\nonumber && - \frac{2 u (\ml^2-2 m_{sp}^2) \ml m_{sp}^2}{(7 \ml^2 - 4 m_{sp}^2)
\left[r - \frac{2aq}{l} + u (3 m_{sp}^2 +4 \ml^2) \right]} \\
\nonumber && - \frac{u ql
  (28 \ml^5-23 \ml^3 m_{sp}^2 +4 \ml m_{sp}^4)}{(7 \ml^2 - 4 m_{sp}^2)
\left[r - \frac{2aq}{l} + u (3 m_{sp}^2 +4 \ml^2) \right]} ,\\
\nonumber B_3 &=&  \frac{4 ql m_{sp}^2}{7 \ml^2-4 m_{sp}^2}
\\
\label{exprB3} && + \frac{l^2 \left(\frac{4a}{l^2} - \frac{4aq}{l} + r - 4 u \ml^2 + u m_{sp}^2\right)}{2a} ,\\
\label{exprE3} E_3 &=& \frac{4 \ml m_{sp}}{7 \ml^2-4 m_{sp}^2} .
\end{eqnarray}
Then we match Eq. (\ref{eta1far}) and Eq. (\ref{eta1near}), and $\eta_3$
by matching Eq. (\ref{eta3far}) and Eq. (\ref{eta3near}) at some distance $\rho
= L$, and we can solve for $l$. Matching is only
possible if $B_3$ and $E_3$ are positive, and $E_1$ is negative. For $E_3$ to be
positive, the denominator $7 \ml^2-4 m_{sp}^2$ has to be
positive. (This makes the Skyrmion gas to appear only if $H \geq 0.798
H_{c2}$.) $L$ depends on $\xi$ slightly. The lack of an
analytic solution forces us to explore the core size numerically.

\begin{table}
\caption{\label{tbl_ACPskyrsol} Isolated Skyrmions in aligned conical
  phase for different values of $\xi_h'$ all for $H = 0.8 H_{c2}$. }
\begin{tabular}{|c|c|c|c|} 
\hline
 $\xi_h'$ & $l$ & $L$ & $\xi=qz-\varphi$\\
\hline
\hline
$3.16 q^{-1}$ ($r=0.7aq^2$) & $1.59 q^{-1}$ & $2.38 q^{-1}$ & $0$ \\
\cline{2-4}
 & $1.59 q^{-1}$ & $2.31 q^{-1}$ & $\frac{\pi}{2}$ \\
\cline{2-4}
 & $1.59 q^{-1}$ & $2.38 q^{-1}$ & $\pi$ \\
\cline{2-4}
 & $1.58 q^{-1}$ & $2.24 q^{-1}$ & $\frac{3\pi}{2}$ \\
\hline
$1.83 q^{-1}$ ($r=0.7aq^2$) & $1.13 q^{-1}$ & $2.13 q^{-1}$ & $0$ \\
\cline{2-4}
 & $1.13 q^{-1}$ & $1.88 q^{-1}$ & $\frac{\pi}{2}$ \\
\cline{2-4}
 & $1.13 q^{-1}$ & $2.13 q^{-1}$ & $\pi$ \\
\cline{2-4}
 & $1.12 q^{-1}$ & $1.38 q^{-1}$ & $\frac{3\pi}{2}$ \\
\hline
$1.41 q^{-1}$ ($r=0.5aq^2$) & $0.88 q^{-1}$ & $1.35 q^{-1}$ & $0$ \\
\cline{2-4}
 & $0.88 q^{-1}$ & $1.29 q^{-1}$ & $\frac{\pi}{2}$ \\
\cline{2-4}
 & $0.88 q^{-1}$ & $1.35 q^{-1}$ & $\pi$ \\
\cline{2-4}
 & $0.88 q^{-1}$ & $1.23 q^{-1}$ & $\frac{3\pi}{2}$ \\
\hline
\hline
$1.29 q^{-1}$ ($r=0.4aq^2$) & $0.79 q^{-1}$ & $1.13 q^{-1}$ & $0$ \\
\cline{2-4}
 & $0.79 q^{-1}$ & $1.15 q^{-1}$ & $\frac{\pi}{2}$ \\
\cline{2-4}
 & $0.79 q^{-1}$ & $1.13 q^{-1}$ & $\pi$ \\
\cline{2-4}
 & $0.80 q^{-1}$ & $1.20 q^{-1}$ & $\frac{3\pi}{2}$ \\
\hline
\end{tabular}
\end{table}

In Table \ref{tbl_ACPskyrsol}, the sizes of Skyrmions in the aligned conical phase
with different values of the correlation length $\xi_h'$ and $H = 0.8
H_{c2}$ are listed. $L$ is the size of a Skyrmion found by
matching method. It depends on the phase angle $\xi$ but it does not vary
significantly. In general, as the system goes away from the phase
boundary (as $\xi_h'$ decreases), the size of the Skyrmion decreases. 

\section{Summary and Conclusion}

In summary, we have studied various aspects of Skyrmions in
helimagnets. We first studied the Skyrmion core, and a
related measure of core size $R$ for paramagnets and ferromagnets. For the
aligned conical phase, this $R$ becomes arbitrary. Extra
perturbations are needed to determine it. We then discussed the
Skyrmion tails. They decay exponentially for
paramagnets and ferromagnets. We determined their decay lengths $l_T$,
which is another measure of the size of Skyrmion. We cannot define $l_T$
for the aligned conical phase because the Skyrmions have algebraic tails, as expected from
Goldstone theorem. We also studied the lattice of Skyrmions. 
Through variational methods, we found that the core-to-core distance in the
lattice is of the order of magnitude $q\xi_p^2$.
This can be compared with the experimentally observed $A$ phase where
the core-to-core distance is found to be $\sim q^{-1}$. Insofar as
$q\xi_p \approx 1$ near the helimagnetic phase boundary, experimental
and theoretical results agree.
Lastly, we introduced the matching radius $L$, as another
measure of the core size. In our studies, the core sizes are of the
order $q^{-1}$ near the helimagnetic phase boundary, consistent with various previous
studies. However, in other parts of the phase diagram, it depends on
other length scales as well such as the thermal correlation
lengths $\xi_p$ or $\xi_f$, and the magnetic length $\barlH$, as shown in
Table \ref{tbl_PMFMskyrsol}. Among all measures of sizes of Skyrmions, we
think that $L$ is the best because it depends on the
whole Skyrmion. We also estimated the size of
Skyrmions in aligned conical phase by matching. It is also of order $q^{-1}$
near the helimagnetic phase boundary, and decreases away from it.

\acknowledgments

This work was supported by the National Science Foundation under
Grants No. DMR-09-29966 and No. DMR-09-01907.

\appendix

\section{Model in $CP^1$ Representation} \label{app_CP1}

To facilitate the study of topologically non-trivial Skyrmions, it
is useful to write the model in terms of the $CP^1$ representation,
which is commonly used in the study of quantum Hall
ferromagnets. \cite{FBCMS97} We first write
\begin{equation}
\bM(\bx) = m(\bx) \bn(\bx) ,
\end{equation}
where $m(\bx)$ and $\bn(\bx)$ denotes the magnitude and direction of
$\bM$. The direction $\bn$ can be written in terms of two-component
spin through the Hopf mapping: \cite{Fra98}
\begin{equation}
\label{hopfmap} n_{\alpha} = z_i^* \sigma_{\alpha}^{ij} z_j ,
\end{equation}
where $\sigma_{\alpha}$ is the Pauli matrix. From now onwards, Greek
indices denote spatial component and Latin indices denote merely
matrix component in spins. The constraint $\bn^2=1$ gives
\begin{equation}
\label{constraintz} z_i^* z_i =1  .
\end{equation}

Because $\bn$ has a definite magnitude, there are only two degrees of
freedom. Therefore, $\bz$ should have two degrees of freedom only,
accomplished by the constraint Eq. (\ref{constraintz}) and fixing the
gauge \cite{HZYPN10}
\begin{equation}
\label{def_gauge} A_{\alpha} = -\frac{i}{2} (z_i^* \partial_{\alpha} z_i -
z_i \partial_{\alpha} z_i^*) ,
\end{equation}
such that the transformation $z_i(\bx) \rightarrow e^{-i \theta(\bx)}
z_i(\bx)$ and $z_i^*(\bx) \rightarrow e^{i \theta(\bx)} z_i^*(\bx)$
does not lead to a change in the physical system.

In this representation, we write the action in Eq. (\ref{expr_action}) as
\begin{eqnarray}
\nonumber && \mathcal{S}[m,\mathbf{A},\bz] \\
\nonumber &=& \int d^3x \left[ \frac{r}{2} m^2 + \frac{u}{4}
m^4 + \frac{a}{2} \partial_{\alpha} m \partial_{\alpha} m - H_{\alpha}
m z_i^* \sigma_{\alpha}^{ij} z_j \right] \\
\nonumber && + 2 am^2 \left[ (\partial_{\alpha} z_i^*) (\partial_{\alpha} z_i) -
  A_{\alpha} A_{\alpha} \right] \\
\label{expr_action_CP1} && + c m^2 \left[z_i^* \sigma_{\alpha}^{ij} z_j A_{\alpha} +
  \frac{i}{2} (\partial_{\alpha} z_i^*) \sigma_{\alpha}^{ij} z_j -
  \frac{i}{2} z_i^* \sigma_{\alpha}^{ij} \partial_{\alpha} z_j \right] .
\end{eqnarray}
The saddle-point equation associated with this representation has to
be done with Lagrangian multiplier because of the constraint in Eq. 
(\ref{constraintz}). Consider $\mathcal{S} + \lambda \int d^3x (z_i^* z_i - 1)$
with $\lambda$ being the Lagrangian multiplier,
the saddle-point equations are 
\begin{widetext}
\begin{equation}
\label{saddleeqn1CP1} (r-3aq^2 - 4 a A_{\alpha} A_{\alpha}) m -
a \partial_{\alpha} \partial_{\alpha} m - 4 a q m n_{\alpha} A_{\alpha} + um^3 - H_{\alpha} n_{\alpha}  
\nonumber + 4 a m
\left[\left(\delta_{ij} \partial_{\alpha} - \frac{iq}{2}
    \sigma_{\alpha}^{ij}\right) z_j^*\right]
\left[\left(\delta_{ik} \partial_{\alpha} + \frac{iq}{2}
  \sigma_{\alpha}^{ik}\right) z_k \right] = 0 ,
\end{equation}
\begin{equation}
\label{saddleeqn2CP1} m \left(\delta_{ij} \partial_{\alpha} - i \delta_{ij} A_{\alpha} +
  \frac{iq}{2} \sigma_{\alpha}^{ij}\right)
\left(\delta_{jk} \partial_{\alpha} - i \delta_{jk} A_{\alpha} +
  \frac{iq}{2} \sigma_{\alpha}^{jk}\right) z_k - i q m \left(n_{\alpha} \partial_{\alpha} z_i + \frac{1}{2}
  z_i \partial_{\alpha} n_{\alpha}\right) + \frac{1}{2a} H_{\alpha}
\sigma_{\alpha}^{ij} z_j = \lambda z_i .
\end{equation}
\end{widetext}

\section{Analytic Expressions for the Paramagnet and the Ferromagnet in
  Mean-field LGW Model} \label{app_PMFM_MFLGW}

For the paramagnet and the ferromagnet, the magnetization can be
solved by the saddle-point equation derived from LGW functional,
\begin{equation}
\label{saddlepteqn_m} rm + um^3 - H = 0 .
\end{equation}
Its analytic solutions are given in the following.

\subsection{Paramagnet}

The paramagnet phase is only valid for $r>0$. Its full analytic expression is given by
\begin{equation}
\label{expr_MP} M_P = \chi_p H f\left(\sqrt{\frac{u}{r^3}} H\right) ,
\end{equation}
where
\begin{equation}
\label{def_f} f(x) = \frac{1}{x} \left[\frac{x}{2} + \sqrt{\frac{1}{27} +
    \left(\frac{x}{2}\right)^2}\right]^{\frac{1}{3}} + \frac{1}{x}
\left[\frac{x}{2} - \sqrt{\frac{1}{27} +
    \left(\frac{x}{2}\right)^2}\right]^{\frac{1}{3}} .
\end{equation}
A Taylor's expansion for small $H$ confirms that $M_p \approx \chi_p
H$ as in Eq. (\ref{PM_smallH}). However, when $r$ approaches 0, $M_p
\approx \left(\frac{H}{u}\right)^{\frac{1}{3}}$, giving the critical
exponent $\delta=3$ for the mean-field theory. \cite{Ma76}

\subsection{Ferromagnet}

The ferromagnet is only for $r<0$. In general the magnetization is
given by
\begin{equation}
\label{expr_MF} M_F = \left\{
\begin{array}{ll} 
M_F^{(0)} \left(\cos \frac{\gamma}{3} + \frac{1}{\sqrt{3}}
  \sin \frac{\gamma}{3} \right) ,  &
\mbox{ for $ H < \frac{2}{\sqrt{27} |r|} \sqrt{\frac{u}{|r|}}$} \\
2 \chi_f H f\left(\sqrt{\frac{u}{|r|^3}} H\right), 
& \mbox{ for $H >
    \frac{2}{\sqrt{27} |r|} \sqrt{\frac{u}{|r|}}$} 
\end{array} 
\right.
\end{equation}
where $M_F^{(0)}$ is given by Eq. (\ref{expr_MFH0}), $f(x)$ is defined in (\ref{def_f}) and 
\begin{equation}
\label{def_gamma} \sin \gamma = \frac{\sqrt{27}}{2 |r|} \sqrt{\frac{u}{|r|}} H .
\end{equation}
A Taylor's expansion for small $H$ confirms that $m \approx M_F^{(0)}
+ \chi_f H$.

\section{Isolated Skyrmion in Non-Linear $\sigma$
  Model} \label{app_skyrNLSM}

In the ferromagnetic phase, the fluctuations can be studied with
non-linear $\sigma$ model.\cite{Tsv03} In this model, there exists a
metastable mean-field solution that corresponds to a Skyrmion given by \cite{AbaPok98}
\begin{eqnarray}
\label{sigmaSkyx} n_x &=& -\frac{2 l \rho \sin\varphi}{l^2+\rho^2} , \\
\label{sigmaSkyy} n_y &=& \frac{2 l \rho \cos\varphi}{l^2+\rho^2} , \\
\label{sigmaSkyz} n_z &=& \frac{\rho^2-l^2}{\rho^2+l^2} .
\end{eqnarray}
where $l$ is an arbitrary length that characterizes the size of the Skyrmion core. In terms of the
representation in Eq. (\ref{defbn}), \cite{LiToBe09}
\begin{eqnarray}
\label{sigmaSkyalpha} \alpha(\bx) &=& \frac{\pi}{2} , \\
\label{sigmaSkytheta} \theta(\bx) &=& 2 \tan^{-1}
\frac{l}{\rho} .
\end{eqnarray}
In $CP^1$ representation,
\begin{equation}
\bz = \sqrt{\frac{\rho^2}{\rho^2+l^2}} \left[ \begin{array}{c}
    e^{i\left(\frac{\pi}{2}-\varphi\right)} \\ \frac{l}{\rho}  \end{array} \right] ,
\end{equation}
which is an anti-vortex. \cite{HZYPN10} This solution has a winding
number $W=-1$, \cite{BelPol75} confirming that it is a Skyrmion.

This Skyrmion in non-linear $\sigma$ model states that
\begin{equation}
\label{SkyrNLSM_approx} \theta(\rho) \approx \left\{ \begin{array}{ll} \pi - \frac{\rho}{2l},
    & \mbox{ for $\rho \approx 0$} \\
\frac{2l}{\rho}, & \mbox{ for $\rho \rightarrow
  \infty$} .
\end{array} \right.
\end{equation} 
We cannot determine $l$ as discussed in Sec. \ref{sec_coreFMH0q0}.

\section{Saddle-Point Equations for $m(\bx)$, $\theta(\bx)$ and $\alpha(\bx)$} 

\subsection{In Field-Polarized Magnets} \label{app_skyrPMFM}

Putting $\bM = m \bn$ with $\bn$ given by Eq. (\ref{defbn}) with
$\alpha(\bx)=\frac{\pi}{2}$ into Eq. (\ref{saddlepteqn}), the
saddle-point equation becomes
\begin{eqnarray}
\nonumber r m - a \left[\left(\frac{\partial^2
      m}{\partial \rho^2} + \frac{1}{\rho} \frac{\partial
      m}{\partial \rho}\right) - m \left(\frac{\partial
      \theta}{\partial \rho}\right)^2 - \frac{1}{\rho^2} m \sin^2
  \theta\right] \\
\nonumber + c m \left(\frac{\partial \theta}{\partial \rho} +
  \frac{1}{\rho} \sin\theta\cos\theta\right) + u m^3 - H\cos\theta =
0 ,\\
\label{saddlepteqnPMFM1} 
\end{eqnarray}
\begin{eqnarray}
\nonumber a m \left( \frac{\partial^2 \theta}{\partial
    \rho^2} + \frac{1}{\rho} \frac{\partial \theta}{\partial \rho} -
  \frac{1}{\rho^2} \sin\theta \cos\theta + \frac{2}{m} \frac{\partial
    m}{\partial \rho} \frac{\partial \theta}{\partial \rho} \right)
 \\
\nonumber + c
\left(\frac{\partial m}{\partial \rho} + \frac{1}{\rho} m \sin^2
  \theta\right) - H \sin\theta = 0 . \\
\label{saddlepteqnPMFM2} 
\end{eqnarray}
These two equations are useful for describing a Skyrmion in
paramagnets and ferromagnets.

\subsection{In helimagnetic phases} \label{app_saddlepteqnHM}

Putting $\bM = m \bn$ with $\bn$ given by Eq. (\ref{defbn}) and with
\begin{equation}
\label{def_xi} \xi = qz-\varphi ,
\end{equation}
the Eq. (\ref{saddlepteqn}) gives the following
equations:
\begin{widetext}
\begin{eqnarray}
\nonumber r m_s \sin\theta \cos\alpha - a \left\{ \frac{1}{\rho}
  \frac{\partial}{\partial \rho} \left[\rho \frac{\partial}{\partial
      \rho} (m_s \sin\theta \cos\alpha)\right] \right\} \\
\nonumber -a \left\{
  \left(\frac{1}{\rho^2}+q^2\right) \frac{\partial^2}{\partial \xi^2}
  (m_s \sin\theta \cos\alpha) + \frac{2}{\rho^2} \frac{\partial}{\partial \xi} (m_s
  \sin\theta \sin\alpha) - \frac{1}{\rho^2} m_s \sin\theta
  \cos\alpha\right\} \\
\label{saddlepteqnHM1} + c \left[- \frac{1}{\rho}
  \frac{\partial}{\partial \xi} (m_s \cos\theta) - q
  \frac{\partial}{\partial \xi} (m_s \sin\theta \sin\alpha)\right] + u m_s^3 \sin\theta \cos\alpha = 0 ,
\end{eqnarray}
\begin{eqnarray}
\nonumber r m_s \sin\theta \sin\alpha - a \left\{ \frac{1}{\rho}
  \frac{\partial}{\partial \rho} \left[\rho \frac{\partial}{\partial
      \rho} (m_s \sin\theta \sin\alpha)\right] \right\} \\
\nonumber -a \left\{
  \left(\frac{1}{\rho^2}+q^2\right) \frac{\partial^2}{\partial \xi^2}
  (m_s \sin\theta \sin\alpha) - \frac{2}{\rho^2} \frac{\partial}{\partial \xi} (m_s
  \sin\theta \cos\alpha) - \frac{1}{\rho^2} m_s \sin\theta
  \sin\alpha\right\} \\
\label{saddlepteqnHM2} + c \left[- \frac{\partial}{\partial \rho} (m_s
\cos\theta) + q \frac{\partial}{\partial \xi} (m_s \sin\theta
\cos\alpha)\right] + u m_s^3 \sin\theta \sin\alpha = 0 ,
\end{eqnarray}
\begin{eqnarray}
\nonumber r m_s \cos\theta - a \left\{ \frac{1}{\rho}
  \frac{\partial}{\partial \rho} \left[\rho \frac{\partial}{\partial
      \rho} (m_s \cos\theta)\right] + \left(\frac{1}{\rho^2}+q^2\right) \frac{\partial^2}{\partial \xi^2}
  (m_s \cos\theta)\right\} \\
\label{saddlepteqnHM3}  + c \left[\frac{1}{\rho}
  \frac{\partial}{\partial \rho} (\rho m_s \sin\theta \sin\alpha) +
  \frac{1}{\rho} \frac{\partial}{\partial \xi} (m_s \sin\theta
  \cos\alpha)\right] + u m_s^3 \cos\theta - H = 0 .
\end{eqnarray}
\end{widetext}

\section{Linear Response Theory} \label{app_LRTSPE}

Linear response theory relates external perturbations of a
physical system to the correlation functions in the unperturbed
one. \cite{For89}  This ensures that the perturbations have long-ranged
behaviors if there exist correlation functions for massless
modes for fluctuations. In contrast, the perturbations are of short-ranged if there do not
exist any massless modes. This explains
the behaviors of Skyrmion tails in different phases in Sec. \ref{sec_tail} in relation to
the Goldstone theorem.

Assume there is a field $\bM(\bx)$ with a known mean-field solution
$\bM_0$, with the corresponding free energy per temperature being
$S_0$. Let the kernel matrix for fluctuations be
$\mathbf{K}(\bx,\bx')$ and an external perturbation $\bH(\bx)$. Then the
corresponding partition function is
\begin{eqnarray}
\nonumber Z &=& e^{-S_0} \int D\bM \cdot \\
\nonumber && \exp\left[- \int d^dx \int d^dx'
  \frac{1}{2} \delta M_i(\bx)
  K_{ij}(\bx,\bx') \delta M_j(\bx') \right. \\
\label{LRTSPEeqn1} && \left. + \beta \int d^dx \cdot H_i(\bx) \delta  M_i(\bx)\right] ,
\end{eqnarray}
where summation is on repeated indices and $\beta = (k_B T)^{-1}$. By
carrying out the functional integral, we get the fluctuation
determinant. \cite{Kle95} The equation
for $\delta M$, by variational principle, is
\begin{equation}
\label{LRTSPEeqn2} \int d^dx' \cdot K_{ij}(\bx,\bx') \delta
M_j(\bx') - \beta H_i(\bx) = 0 ,
\end{equation}
and hence,
\begin{equation}
\label{LRTSPEeqn3} \delta M_i(\bx) = \beta \int d^dx' \cdot
K_{ij}^{-1}(\bx,\bx') H_j(\bx') .
\end{equation}
Note that the inverse of the kernel matrix is actually the correlation matrix. 
We can see from here that the perturbation $\delta \bM(\bx)$ has the same
behavior as the correlation functions $\mathbf{K} (\bx,\bx')$. If
$K^{-1}(\bx,\bx') = K^{-1}(\bx-\bx')$, the Fourier representation of
Eq. (\ref{LRTSPEeqn3}) is
\begin{equation}
\label{LRTSPEeqn4} \delta M_i(\bk) = \beta K_{ij}^{-1}(\bk) H_j(\bk) .
\end{equation}

We illustrate this with the example of a pure ferromagnet with $q=0$
and $H=0$. Assume that the magnet is aligned along $z$ direction. Then the
kernel matrices (the inverse of the correlation functions) are given by \cite{Ma76}
\begin{eqnarray*}
K_{ij} (\bk) = \delta_{ij} (a k^2 - 2 r \delta_{j z}) .
\end{eqnarray*}
This indicates that along $x$ and $y$ directions, the fluctuations are
massless as expected by Goldstone theorem. Along $z$
direction, it is massive. Then, we get
\begin{eqnarray*}
a k^2 \delta M_x(\bk) - \beta H_x(\bk) = 0, \\
a k^2 \delta M_y(\bk) - \beta H_y(\bk) = 0, \\
(a k^2-2r) \delta M_z(\bk) - \beta H_z(\bk) = 0.
\end{eqnarray*}
This gives $\delta M_x (\bx), \delta M_y(\bx) \sim |\bx|^{-1}$ and
$\delta M_z(\bx) \sim |\bx|^{-1} e^{-\frac{|\bx|}{\xi_f}}$. Here, we have demonstrated that the correlation
function of a massless mode leads to long-range behaviors of the
perturbations. Similar behavior exists in the perturbations to aligned
conical phase in Sec. \ref{sec_tailACP}.

Here, we illustrate an example of a pure paramagnet with $q=0$ and $H=0$
here. In this case, $\bM_0 = 0$. The kernel matrices are \cite{Ma76}
\begin{eqnarray*}
K_{ij} (\bk) = \delta_{ij} (ak^2 -r) .
\end{eqnarray*}
Then
\begin{eqnarray*}
(a k^2-r) \delta M_i(\bk) - \beta H_i(\bk) = 0,
\end{eqnarray*}
for $i=x,y$ and $z$.
This gives $\delta \bM(\bx) \sim |\bx|^{-1}
e^{-\frac{|\bx|}{\xi_p}}$. Here we have demonstrated that if the correlation functions of all modes
are massive, then the perturbations are short-ranged.


\bibliography{helimagnet}

\end{document}